\documentstyle[twoside]{article}
\catcode`\@=11
\long\def\@makefntext#1{ 
\protect\noindent \hbox to 3.2pt {\hskip-.9pt
$^{{\eightrm\@thefnmark}}$\hfil}#1\hfill} 

\def\thefootnote{\fnsymbol{footnote}}
 \def\@makefnmark{\hbox to 0pt{$^{\@thefnmark}$\hss}}  

\def\ps@myheadings{\let\@mkboth\@gobbletwo
\def\@oddhead{\hbox{} 
\rightmark\hfil\eightrm\thepage}
\def\@oddfoot{}\def\@evenhead{\eightrm\thepage\hfil 
\leftmark\hbox{}}\def\@evenfoot{}
\def\sectionmark##1{}\def\subsectionmark##1{}}

\oddsidemargin=\evensidemargin
\renewcommand{\thefootnote}{\fnsymbol{footnote}}

\newcounter{sectionc}\newcounter{subsectionc}\newcounter{subsubsectionc}
\renewcommand{\section}[1] {\vspace{12pt}\addtocounter{sectionc}{1}
\setcounter{subsectionc}{0}\setcounter{subsubsectionc}{0}\noindent
	{\tenbf\thesectionc. #1}\par\vspace{5pt}}
\renewcommand{\subsection}[1] {\vspace{12pt}\addtocounter{subsectionc}{1}
	\setcounter{subsubsectionc}{0}\noindent
	{\bf\thesectionc.\thesubsectionc. {\kern1pt \bfit #1}}\par\vspace{5pt}}
\renewcommand{\subsubsection}[1] {\vspace{12pt}
\addtocounter{subsubsectionc}{1}
	\noindent{\tenrm\thesectionc.\thesubsectionc.\thesubsubsectionc.
	{\kern1pt \tenit #1}}\par\vspace{5pt}}
\newcommand{\nonumsection}[1] {\vspace{12pt}\noindent{\tenbf #1}
	\par\vspace{5pt}}

\newcounter{appendixc}
\newcounter{subappendixc}[appendixc]
\newcounter{subsubappendixc}[subappendixc]
\renewcommand{\thesubappendixc}{\Alph{appendixc}.\arabic{subappendixc}}
\renewcommand{\thesubsubappendixc}
	{\Alph{appendixc}.\arabic{subappendixc}.\arabic{subsubappendixc}}

\renewcommand{\appendix}[1] {\vspace{12pt}
        \refstepcounter{appendixc}
        \setcounter{figure}{0}
        \setcounter{table}{0}
        \setcounter{lemma}{0}
        \setcounter{theorem}{0}
        \setcounter{corollary}{0}
        \setcounter{definition}{0}
        \setcounter{equation}{0}
        \renewcommand{\thefigure}{\Alph{appendixc}.\arabic{figure}}
        \renewcommand{\thetable}{\Alph{appendixc}.\arabic{table}}
        \renewcommand{\theappendixc}{\Alph{appendixc}}
        \renewcommand{\thelemma}{\Alph{appendixc}.\arabic{lemma}}
        \renewcommand{\thetheorem}{\Alph{appendixc}.\arabic{theorem}}
        \renewcommand{\thedefinition}{\Alph{appendixc}.\arabic{definition}}
        \renewcommand{\thecorollary}{\Alph{appendixc}.\arabic{corollary}}
        \renewcommand{\theequation}{\Alph{appendixc}.\arabic{equation}}
        \noindent{\tenbf Appendix \theappendixc #1}\par\vspace{5pt}}
\newcommand{\subappendix}[1] {\vspace{12pt}
        \refstepcounter{subappendixc}
        \noindent{\bf Appendix \thesubappendixc. {\kern1pt \bfit #1}}
	\par\vspace{5pt}}
\newcommand{\subsubappendix}[1] {\vspace{12pt}
        \refstepcounter{subsubappendixc}
        \noindent{\rm Appendix \thesubsubappendixc. {\kern1pt \tenit #1}}
	\par\vspace{5pt}}

\newcommand{\textlineskip}{\baselineskip=13pt}
\newcommand{\smalllineskip}{\baselineskip=10pt}

\def\eightcirc{
\begin{picture}(0,0)
\put(4.4,1.8){\circle{6.5}}
\end{picture}}
\def\eightcopyright{\eightcirc\kern2.7pt\hbox{\eightrm c}}

\newcommand{\copyrightheading}[1]
	{\vspace*{-2.5cm}\smalllineskip{\flushleft
	{\eightrm $\phantom{International Journal of Modern Physics D, #1}$
}\\
	{\eightrm $\phantom{, World Scientific Publishing
 Company}$}\\
	 }}

\def\abstracts#1#2#3{{
	\centering{\begin{minipage}{4.5in}\baselineskip=10pt\eightrm
	\centerline{ABSTRACT}
	\parindent=0pt #1\par
	\parindent=15pt #2\par
	\parindent=15pt #3
	\end{minipage} }\par}}

\newcommand{\bibit}{\nineit}

\renewenvironment{thebibliography}[1]			
	{\ninerm
	 \baselineskip=11pt				
	 \begin{list}{\arabic{enumi}.}
	{\usecounter{enumi}\setlength{\parsep}{0pt}
	 \setlength{\leftmargin 17pt}{\rightmargin 0pt}	
	 \setlength{\itemsep}{0pt} \settowidth		
	{\labelwidth}{#1.}\sloppy}}{\end{list}}

\newcounter{itemlistc}
\newcounter{romanlistc}
\newcounter{alphlistc}
\newcounter{arabiclistc}

\newcommand{\fcaption}[1]{
        \refstepcounter{figure}
        \setbox\@tempboxa = \hbox{\eightrm Fig.~\thefigure. #1}
        \ifdim \wd\@tempboxa > 5in
           {\begin{center}
        \parbox{5in}{\eightrm \smalllineskip Fig.~\thefigure. #1 }
            \end{center}}
        \else
             {\begin{center}
             {\eightrm Fig.~\thefigure. #1}
              \end{center}}
        \fi}

\newcommand{\tcaption}[1]{
        \refstepcounter{table}
        \setbox\@tempboxa = \hbox{\eightrm Table~\thetable. #1}
        \ifdim \wd\@tempboxa > 5in
           {\begin{center}
        \parbox{5in}{\eightrm\smalllineskip Table~\thetable. #1 }
            \end{center}}
        \else
             {\begin{center}
             {\eightrm Table~\thetable. #1}
              \end{center}}
        \fi}

\def\@citex[#1]#2{\if@filesw\immediate\write\@auxout	
	{\string\citation{#2}}\fi			
\def\@citea{}\@cite{\@for\@citeb:=#2\do			
	{\@citea\def\@citea{,}\@ifundefined		
	{b@\@citeb}{{\bf ?}\@warning
	{Citation `\@citeb' on page \thepage \space undefined}}
	{\csname b@\@citeb\endcsname}}}{#1}}

\newif\if@cghi
\def\cite{\@cghitrue\@ifnextchar [{\@tempswatrue
	\@citex}{\@tempswafalse\@citex[]}}
\def\citelow{\@cghifalse\@ifnextchar [{\@tempswatrue
	\@citex}{\@tempswafalse\@citex[]}}
\def\@cite#1#2{{$\null^{#1}$\if@tempswa\typeout
	{IJCGA warning: optional citation argument
	ignored: `#2'} \fi}}

\def\pmb#1{\setbox0=\hbox{#1}
	\kern-.025em\copy0\kern-\wd0
	\kern.05em\copy0\kern-\wd0
	\kern-.025em\raise.0433em\box0}

\def\fnm#1{$^{\mbox{\scriptsize #1}}$}
\def\fnt#1#2{\footnotetext{\kern-.3em
	{$^{\mbox{\scriptsize #1}}$}{#2}}}

\def\fpage#1{\begingroup
\voffset=.3in
\thispagestyle{empty}\begin{table}[b]\centerline{\footnotesize #1}
	\end{table}\endgroup}

\font\tenbf=cmbx10
\font\tenit=cmti10
\font\tenit=cmti10
\font\bfit=cmbxti10 at 10pt
 1
 1
 1

\font\ninerm=cmr9
\font\nineit=cmti9

\font\eightrm=cmr8

\def\qed{\hbox{${\vcenter{\vbox{
   \hrule height 0.4pt\hbox{\vrule width 0.4pt height 6pt
   \kern5pt\vrule width 0.4pt}\hrule height 0.4pt}}}$}}

\begin{document}
\normalsize\textlineskip
{\thispagestyle{empty}
\setcounter{page}{1}

\renewcommand{\thefootnote}{\fnsymbol{footnote}}

\copyrightheading{}
\begin{flushright}
ULB-TH-06/93 \\ June 1993
\end{flushright}
\vspace*{0.88truein}

\fpage{$\phantom{1}$}
\centerline{\bf MATRIX MODELS AND INTEGRABLE C $<$ 1}
\vspace*{0.035truein}
\centerline{\bf OPEN STRING THEORIES\footnote{Talk presented
at the Journ\'ees Relativistes '93, Brussels, April, 1993.}}
\vspace{0.37truein}
\centerline{\footnotesize LAURENT HOUART\footnote{Aspirant FNRS,
e-mail: lhouart@ulb.ac.be}}
\vspace*{0.015truein}
\centerline{\footnotesize\it Service de Physique Th\'eorique,
Universit\'e Libre de Bruxelles, Boulevard du Triomphe}
\baselineskip=10pt
\centerline{\footnotesize\it CP 225, B-1050 Bruxelles, Belgium}
\vglue 10pt
\vspace*{0.015truein}

\baselineskip=10pt
\vspace{0.225truein}

\vspace*{0.21truein}
\abstracts{\noindent  We study in the double scaling limit the
two-matrix model which represents the sum over closed and open
random surfaces coupled to an Ising model.
The boundary conditions are characterized by the fact that
the Ising spins sitting at the vertices of the boundaries
are all in the same state. We obtain the string equation.}{}{}

\vspace*{-3pt}\textlineskip
\textheight=7.8truein
\setcounter{footnote}{0}
\renewcommand{\thefootnote}{\alph{footnote}}

\nonumsection{}

Since the discovery of the double scaling limit\cite{brez},
important progress has been achieved in the understanding of
open-closed string theories in $(p,q)$ conformal minimal model
backgrounds.

We have now a complete coherent description of the $(2m-1,2)$
conformal minimal models coupled to 2D gravity. These theories are
described by the KdV hierarchy associated with the lie algebra
$sl(2,C)$. Starting with the usual one-hermitian-matrix
model\cite{brez}, which describes the sum over closed
surfaces, one finds that  the KdV flow equations organize
the operator structure of the $(2m-1,2)$ theories. Supplying
``initial conditions"
given by the string equation yields a complete description of
closed string theories in the $(2m-1,2)$ backgrounds.\hfill \break
The operator content of the theories is fully understood.
Indeed, the infinite number of operators found by BRST analysis
in the Liouville description\cite{zuck}is also present in the KdV
formulation. The remaining KdV flows are identified studying the
macroscopic loops\cite{bank,mart,stau}. They correspond to
boundary operators\cite{mart} (the boundary length and the infinite
number of operators associated with it). The KdV hierarchy
associated
to $sl(2,C)$ is thus the good framework to describe both closed and
open string theories in $(2m-1,2)$ conformal minimal
backgrounds.\hfill
\break  Moreover, a direct study of the one-hermitian-matrix model
supplemented with a logarithmic potential\cite{akaz}-- having the
effect of adding surfaces with boundaries of finite extent in the
partition function-- has been performed in the double scaling
limit\cite{kost}. This leads to the generalization of the previous
string equation to the case of a non-zero open string
coupling.\hfill
\break Finally, it has been shown\cite{watt} that the
$sl(2,C)$ mKdV models are a different description of the
same open-closed string theories.

The description of the general (p,q)
models\fnm{a}\fnt{a}{ We consider $p>q$ .} coupled to 2D gravity
is much more involved. Closed string
theories in these backgrounds have been studied. They are
realized in terms of multi-matrix models and are characterized
in the double scaling limit by the generalized KdV hierarchies
\cite{doug,goul,difr}. For a given $q$ the $sl(q,C)$
KdV hierarchy organizes the operator structure of
the theory. Once more,
the spectrum of the theory contains the operators found in the
Liouville description but some remaining flows no longer
admit an interpretation in terms of boundary operators\cite{mart}.
Furthermore, the $sl(q,C)$ KdV hierarchy does not allow a
description of the boundary length.\hfill \break
Related to this problem is the issue of finding a string
equation describing open string theories in $(p,q)$ backgrounds.
In this case, when surfaces with boundaries of finite
extent are added to the partition sum, one has to pay attention
to the boundary problem. There are indeed $q-1$ order
parameters in the theory and the boundary conditions associated
with them have to be fixed. \hfill \break
Recently a string equation for general open string theories in
$(p,q)$ backgrounds has been derived by Johnson \cite{john} in the
integrable model framework. This equation is obtained by studying
the $sl(q,C)$ generalization of a mapping which transforms
the $sl(2,C)$ $\tau$-function characterizing the closed case
into the $sl(2,C)$ $\tau$-function of the open case. However
by making use of the beautiful underlying mathematical
structure and bypassing the matrix model route some physical
informations are lost. It is very hard to know what are the
boundary conditions associated with the string equation.

This talk is concerned with the boundary condition
problem. Starting with the discrete formulation, I work out an
explicit example: the open string theory in the $(4,3)$ conformal
background with fixed boundary conditions. In the spirit of
ref.[7], using the orthogonal polynomial method, I perform
the double scaling limit on the  two-matrix model which
represents the sum over closed and open  random surfaces coupled
to an Ising model (IM). The boundary conditions are characterized
by the fact that the Ising spins sitting at the vertices
of the boundaries are all in the same state. Here I present
only the results and refer the reader to ref.[20] for more details.

The model is given by the following two-matrix model: $F=ln Z$
\begin{equation}
Z=\int {\cal D}M_1 {\cal D}M_2 \exp{(-{N \over g}
TrV(M_1,M_2))} \label{eqi}
\end{equation}

where $V$ is the asymmetrical potential:
\begin{equation}
V(M_1,M_2)={1 \over 2} M^2_1 -{1 \over 4} M^4_1 +
{1 \over 2} M^2_2 -{1 \over 4} M^4_2 - cM_1 M_2 + \gamma
\ln(1-e^{\mu}M^2_1)\label{eqii}
\end{equation}
When $\gamma =0$, $F$ corresponds to the sum over closed surfaces
coupled to an Ising model\cite{kaza}. The two $N \times N$
hermitian matrices $M_1,M_2$
are used to represent the two spin states. The parameter $g$ is
related to the bulk cosmological constant and $N$ is the closed
string coupling. The solution of the model
in this case is well-known\cite{crmo,isin,migd}.

\noindent
When $\gamma \not= 0$ the logarithmic potential has the effect
of adding surfaces with arbitrary boundaries of finite
extent to the partition function\cite{akaz,kost}. In this case,
$F$ represents the sum over closed and open triangulated
surfaces with the Ising spins sitting at the vertices of the
boundaries all in the same state (e.g. all up). The parameter
$\gamma$ is the open string coupling. In the continuum limit $\mu$
will correspond to the parameter which couples to the boundary
length as well as the parameter which couples to the
boundary magnetization (there is no way to distinguish them in
this model).

The model can be solved non-perturbatively using the well-known
orthogonal polynomial method\cite{itz}. The string equation
found in the double scaling limit is given by:
\begin{equation}
 G^2 w^{\prime \prime}-3uw = 27  \Gamma G R_0(z,M) \label{eqiii}
\end{equation}
\begin{eqnarray}
& & {6 \over 5} \lbrack {1 \over 3} u^3+{8 \over
(12)^2}G^4 u^{(4)}-{1 \over 2}G^2 u^{\prime \prime}u-
{1 \over 4}G^2 (u^\prime)^2+w^2 \rbrack  =  z \nonumber \\
& & +{54 \over 5}G^2
\Gamma \lbrack 2R_1(z,M)-R^\prime_0(z,M) \rbrack \label{eqiv}
\end{eqnarray}
where $z$ is the continuum bulk cosmological constant, $G$ is the
renormalized closed string coupling, $\Gamma$ is the
renormalized open
string coupling and ${d^2 \over dz^2}F=-{1 \over G^2}u(z)$.
The $Z_2$ breaking field $w$ is proportional to
$\langle 1 \sigma \rangle$.
$R_0(z,M)$ is the resolvent and $R_1(z,M)$ is the first jet of the
resolvent\cite{gelf} associated with the operator
$L=G^3 \partial^3_z -{3 \over 4}G \lbrace u,\partial
\rbrace + {3 \over 2}w$ , to wit:
\begin{equation}
R_0(z,M) \equiv \langle z \mid {1 \over M+L} \mid
z \rangle \label{eqv}
\end{equation}
\begin{equation}
R_1(z,M) \equiv \partial_{\tilde z} \langle z \mid {1
\over M+L} \mid \tilde z \rangle \biggl
\vert_{\tilde z= z}\label{eqvi}
\end{equation}
The constant $M$, which comes from the bare parameter $\mu$, is
the renormalized boundary magnetization constant.

Now, using the Adler mapping\cite{gelf},
we compute a system of two equations for
the resolvent $R_0$ and the first jet $R_1$. The result is:
\begin{eqnarray}
 & & - G^3  (2R_1-R^\prime_0)^{(3)}+{3 \over 4}G u^\prime
(2R_1-R^\prime_0) + {3 \over 2}G
u(2R_1-R^\prime_0)^\prime \nonumber \\
& &  = 3 w^\prime R_0 +{9 \over 2}wR^\prime_0 +3 M R^\prime_0
\label{eqvii}
\end{eqnarray}
\begin{eqnarray}
 & &  {1 \over 9}G^4 R^{(5)}_0 -{5 \over 6} G^2
 u R^{(3)}_0- {5 \over 4}G^2u^\prime R^{(2)}_0-({3 \over 4}
G^2 u^{\prime \prime}  - u^2) R^\prime_0   - ({1 \over 6}
G^2 u^{(3)}  -u u^\prime) R_0 \nonumber \\
& & = {1 \over 2}G
w^\prime (2R_1-R^\prime_0) +{3 \over 2} G w
(2R_1-R^\prime_0)^\prime+G M (2R_1-R^\prime_0)^\prime
\label{eqviii}
\end{eqnarray}

We then multiply eq.(\ref{eqvii}) by $G(2R_1-R^\prime_0)$ and
add to it eq.(\ref{eqviii}) multiplied by $3R_0$. It gives us a
total differential which integrated once with respect to $z$
leads to:
\begin{eqnarray}
& & {3 \over 4}G^2u T^2-G^4 \lbrack TT^{(2)}-{1 \over 2}
(T^\prime)^2 \rbrack +{1 \over 3}G^4 \lbrack R_0R^{(4)}_0-
R^\prime_0 R^{(3)}_0 \rbrack \nonumber \\
& & +{1 \over 6}G^4 (R^{\prime \prime}_0)^2-{5 \over 2}G^2u
R_0R^{(2)}_0-{5 \over 4}G^2 \lbrack u^\prime R_0R^\prime_0 - u
(R^\prime_0)^2 \rbrack \nonumber \\
& & -{1 \over 2}G^2u^{(2)}R^2_0 + {3 \over 2}u^2R^2_0-3GTR_0
(M+ {3 \over 2}w)={2 \over 3} G^{-2}
\label{eqix}
\end{eqnarray}
where $T\equiv 2R_1-R^\prime_0$.

\noindent
We have fixed the integration constant using the fact that
$R_0$($u$=$w$=0) $={1 \over 3}G^{-1}M^{-{2 \over 3}}$
and $R_1$($u$=$w$=0) $=-{1 \over 3}G^{-2}M^{-{1 \over 3}}$ .

The model (\ref{eqi}) is thus completely solved in the double
scaling
limit. The solution is given by eqs.(\ref{eqiii}), (\ref{eqiv})
and (\ref{eqix}).
The results obtained here are compatible with those of ref.[12].
They confirm also that the $sl(3,C) \; mKdV$ models are
an equivalent description of open string theories in (p,3)
backgrounds\cite{hou}.

\nonumsection{References}


\noindent
\begin{thebibliography}{000}
\bibitem{brez} E. Br\'ezin and V. Kazakov, {\bibit Phys. Lett.}
{\bf B236} (1990) 144; M. Douglas and S. Shenker,
{\bibit Nucl. Phys.} {\bf B335} (1990) 635; D. Gross and
A.A. Migdal, {\bibit Phys. Rev. Lett.} {\bf 64} (1990) 127;
{\bibit Nucl. Phys.} {\bf B340} (1990) 333.
\bibitem{zuck} B. Lian and G. Zuckerman
{\bibit Phys. Lett.} {\bf B254} (1991) 417.
\bibitem{bank} T. Banks, M. Douglas, N. Seiberg and S. Shenker,
{\bibit Phys. Lett.} {\bf B238} (1990) 279.
\bibitem{mart} E. Martinec, G. Moore and N. Seiberg,
{\bibit Phys. Lett.} {\bf B263} (1991) 190.
\bibitem{stau} G. Moore, N. Seiberg and M. Staudacher,
{\bibit Nucl. Phys.} {\bf B362} (1991) 665.
\bibitem{akaz} V.A. Kazakov, {\bibit Phys. Lett.} {\bf B237}
(1990) 212.
\bibitem{kost} I.K. Kostov, {\bibit Phys. Lett.} {\bf B238}
(1990) 181.
\bibitem{watt} S. Dalley, C.V. Johnson, T.R. Morris and A.
W\"atterstam, {\bibit Mod. Phys. Lett.} {\bf A7} (1992) 2753.
\bibitem{doug} M. Douglas, {\bibit Phys. Lett.} {\bf B238}
(1990) 176.
\bibitem{goul} P. Ginsparg, M. Goulian, M.R. Plesser and J.
Zinn-Justin, {\bibit Nucl. Phys.} {\bf B342} (1990) 539.
\bibitem{difr} P. Di Francesco and D. Kutasov, in
{\bibit proceedings of the Cargese Workshop "Random Surfaces and
Quantum Gravity"}, ed. O. Alvarez
{\bibit et al.} (Plenum Press, New York 1991).
\bibitem{john} C.V. Johnson, Princeton preprint IASSNS-HEP-93/5
(hep-th 9301112).
\bibitem{kaza} V.A. Kazakov,
{\bibit Phys. Lett.} {\bf A119} (1986) 140; D.V. Boulatov and
V.A. Kazakov, {\bibit Phys. Lett.} {\bf B186} (1987) 379.
\bibitem{crmo} \u C. Crnkovi\'c, P. Ginsparg and G. Moore,
{\bibit Phys. Lett.} {\bf B237} (1990) 196.
\bibitem{isin} E. Br\'ezin, M. Douglas, V. Kazakov and S. Shenker,
{\bibit Phys. Lett.} {\bf B237} (1990) 43.
\bibitem{migd} D. Gross and A.A. Migdal,
{\bibit Phys. Rev. Lett.} {\bf 64} (1990) 717.
\bibitem{itz} C. Itzykson and J.-B. Zuber, {\bibit J. Math. Phys.}
{\bf 21} (1980) 411; D. Bessis, C. Itzykson and J.-B. Zuber,
{\bibit Adv. Appl. Math.} {\bf 1} (1980) 109.
\bibitem{gelf} L.A. Dickey, {\bibit Soliton Equations and
Hamiltonian Systems} (World Scientific, Singapore, 1991), pp. 17-20.
\bibitem{hou} L. Houart, {\bibit Phys. Lett.} {\bf B295} (1992) 37.
\bibitem{lho} L. Houart, preprint ULB-TH-02/93 (hep-th 9303157),
to appear in {\bibit Phys. Lett.} {\bf B}.
\end{thebibliography}
\end{document}